\documentclass{article}
\pdfoutput=1
\usepackage{graphicx} 
\usepackage{amsmath}
\usepackage{amsthm}
\usepackage{amsfonts}

\usepackage{amsmath, amsthm, amsfonts, amssymb, bbm}
\usepackage[mathscr]{eucal}

\theoremstyle{plain}
\newtheorem{proposition}{Proposition}

\theoremstyle{definition}

\newtheorem{example}[proposition]{Example}

\theoremstyle{remark}
\newtheorem{remark}[proposition]{Remark}

\newcommand{\ud}{\mathrm{d}}
\newcommand{\del}{\partial}

\newcommand{\skal}[2]{\langle #1 , #2 \rangle}

\newcommand{\R}{\mathbb{R}}

\newcommand{\Q}{\mathcal{Q}}

\newcommand{\order}{\mathcal{O}}

\newcommand{\rhs}{r.h.s.\ }

\newcommand{\wrt}{w.r.t.\ }

\title{Utility based pricing and hedging of jump diffusion processes with a view to applications}
\author{Jochen Zahn \\ Courant Research Centre ``Higher Order Structures'' \\ University of G\"ottingen \\ Bunsenstra{\ss}e 3-5, D-37073 G\"ottingen, Germany \\ jzahn@uni-math.gwdg.de}

\begin{document}

\maketitle

\begin{abstract}
We discuss utility based pricing and hedging of jump diffusion processes with emphasis on the practical applicability of the framework. We point out two difficulties that seem to limit this applicability, namely drift dependence and essential risk aversion independence. We suggest to solve these by a re-interpretation of the framework. This leads to the notion of an implied drift. We also present a heuristic derivation of the marginal indifference price and the marginal optimal hedge that might be useful in numerical computations.
\end{abstract}

\section{Introduction}

The applicability of the Black--Scholes framework for the pricing and hedging of derivative claims crucially depends on the assumption of market completeness, i.e., the possibility to replicate claims and thus eliminate risk. This assumption is not fulfilled if the asset process is driven by more than one source of risk or when market imperfections such as transaction costs are not negligible. One then speaks of an incomplete market in which investors may attribute different prices to derivatives, according to their risk preferences.

As an example, let us consider a jump diffusion process, i.e., the asset $S$ evolves according to 
\begin{equation}
\label{eq:AssetProcess}
	\ud S_t = \mu S_{t_-} \ud t + \sigma S_{t_-} \ud W_t + (e^{J_t} - 1) S_{t_-} \ud N_t. 
\end{equation}
Here $W_t$ is a Wiener and $N_t$ a Poisson process with frequency $\lambda$.
The random variable $J_t$ determines the relative size $e^{J_t}-1$ of the jump. The oldest and probably most popular approach for the pricing and hedging of a claim on such an asset is Merton's \cite{Merton}. There, the investor sets up a portfolio $\Pi$ consisting of the claim with value $V$ and a quantity $- \Delta$ of assets, so that the evolution of the portfolio is given by 
\begin{align}
	\ud \Pi_t = & \ \ud V_t - \Delta_t \ud S_t \nonumber \\
\label{eq:Portfolio}
						= & \left( \del_t V_t(S_t) + \mu S_{t} \del_S V_t(S_{t}) + \tfrac{\sigma^2}{2} S^2_{t} \del_S^2 V_t(S_{t}) - \Delta_t \mu S_{t} \right) \ud t \\
							& + \sigma S_{t} \left\{ \del_S V_t(S_{t}) - \Delta_t \right\} \ud W_t  + \left\{ V_t(e^{J_t} S_{t}) - V_t(S_{t}) - \Delta_t (e^{J_t}-1) S_{t} \right\} \ud N_t \nonumber
\end{align}
It is in general not possible to eliminate jump and diffusion risk at the same time, so some ``optimal'' choice is necessary. Merton's proposal is to hedge only the diffusion risk and to diversify the jump risk, i.e., to set $\Delta_t = \del_S V_t$. The above then yields
\begin{align*}
	\ud \Pi_t = & \left( \del_t V_t(S_t) + \tfrac{\sigma^2}{2} S^2_{t} \del_S^2 V_t(S_{t})  \right) \ud t  \\
							& + \left\{ V_t(e^{J_t} S_{t}) - V_t(S_{t}) - (e^{J_t}-1) S_{t} \del_S V_t(S_t)\right\} \ud N_t.
\end{align*}
If jump risk is diversified, the investor does not need any risk premium for taking this risk, i.e., the expected value of $\ud \Pi_t$ should vanish. Thus, we obtain the partial integro-differential equation (PIDE)
\begin{multline}
\label{eq:MertonPIDE}
0 = \del_t V_t(S) + \tfrac{\sigma^2}{2} S^2 \del_S^2 V_t(S) - \left\{ \int (e^z-1) \ud \nu(z) \right\} S \del_S V_t(S) 
						\\ + \int \left\{ V_t(e^z S) - V_t(S) \right\} \ud \nu(z). 
\end{multline}
Here $\nu$ is the cumulative jump frequency distribution, i.e., for an interval $I$ with characteristic function $\chi_I$, $\nu(\chi_I)$ gives the frequency of jumps of size in $I$. In particular $\nu(\R) = \lambda$. 

Two remarks are in order here:
\begin{enumerate}
  \item The diversification of jump risk is problematic not only in our model (as there is only one asset), but also in practice: In a typical market crash, jumps occur in the whole market, so diversification may well turn out to be accumulation of risk.
  \item Merton's proposal coincides with a naive interpretation of the Black--Scholes framework which states that for risk-neutral pricing one simply has to adjust the drift term such that the expected drift vanishes (in discounted units), and that the appropriate hedging strategy is given by the derivative of the price. In particular, the real-world drift does not enter the price, which is a benefit, as it is notoriously hard to estimate.
\end{enumerate}

Note that the assumption that diversification is possible is crucial here, since otherwise one could not invoke no-arbitrage arguments to set the expected return of the portfolio to zero. If one drops this assumption, then the investor should (i) try to find an optimal balance between diffusion and jump risk and (ii) value the remaining risk in order to obtain a risk premium. A popular framework that achieves (i) is minimal variance pricing and hedging, cf. \cite{Schweizer94,GourierouxEtAl98} and references therein. There, the investor tries to minimize the variance of the expected returns. It has the advantage that no new concepts have to be introduced. However, the choice of a quadratic criterion is somewhat arbitrary and penalizes profits as well as losses.
Furthermore, in the case of a jump diffusion, the framework in general yields a \emph{signed} risk-neutral measure, i.e., there would be positive claims which have a negative value in the framework.
Finally, the framework only tackles (i), but does not yield a price for the remaining risk.

A framework that achieves (i) and (ii) at one stroke is utility based pricing and hedging. There, the investor is equipped with a concave von Neumann utility function $U(X_T)$ that assigns an economic value to the wealth $X_T$ at the investment horizon $T$. Risk aversion is encoded in the concavity of $U$ which entails that the investor prefers a secure income to a random income with the same expectation. This preference is encoded in the risk aversion $A(x) = - U''(x)/U'(x)$.

In this framework, the appropriate price $v$ for a claim with payoff $C(S_T)$ is the indifference price, i.e., the amount the investor should receive such that her maximal expected utility $E[U(X_T - C(S_T))]$ for initial capital $x + v$ is the same as the expected utility $E[U(X_T)]$ for initial capital $x$. This means that one has to consider investment and hedging at the same time and then try to disentangle them. In general, this is a very complicated optimization problem. However, in the limit where the number of traded claims is infinitesimally small, the problem becomes much simpler. One then speaks of the \emph{marginal indifference price} and the corresponding \emph{marginal optimal hedge}. This field has ripen considerably during the last years. Milestones were the papers of Kramkov and S{\^\i}rbu, who gave sufficient criteria for the the marginal indifference price to be well-behaved \cite{KS06}, and defined the concept of the marginal optimal hedge, together with convenient characterizations of it \cite{KS07}. The framework was applied to, e.g., basis risk \cite{KS07}, transaction costs \cite{Wilmott, Monoyios}, and L\'evy processes and stochastic volatility models \cite{KMV09}.

In spite of its conceptual elegance, the practical applicability of the framework seems to be limited by two problems:
\begin{enumerate}
 \item The marginal indifference price and the marginal optimal hedge depend strongly on the real-world drift, which is notoriously hard to estimate.
 \item The marginal indifference price (and also the marginal optimal hedge) are essentially independent of the risk aversion of the investor \cite{Nutz10}.
\end{enumerate}
The second fact is quite inconvenient for a framework whose purpose is to incorporate risk aversion. It turns out that the two problems can be solved, at one stroke, by a change of perspective. The marginal indifference price takes into account how well the option trade matches to the optimal investment strategy of the investor. The assumption is of course that the investor is invested in this optimal strategy. However, the investment strategy a bank chooses is typically not derived from the model that is used to price options. It is thus tempting to interpret the \emph{actual} investment strategy as the optimal one and adjust the drift such that they match. One may thus speak of an \emph{implied drift}. To the best of our knowledge, this concept is new.

In the following section, we introduce utility based pricing and hedging in a heuristic fashion. In particular, we do not (explicitly) use semi-martingale decompositions, on which the rigorous mathematical treatment \cite{KS06, KS07} heavily relies. 
Instead, we use the concept of functional differentiation and derive a formula for the marginal optimal hedge that is similar to the well-known $\Delta$ hedging formula. It has a straightforward economic interpretation and is, to the best of our knowledge, new. We suspect that, when made mathematically precise, this approach is equivalent to setting of Kramkov and S{\^\i}rbu in the common domain of applicability. At least for the case of a jump diffusion under power utility, this is shown to be true, as we re-derive results implicitly contained \cite{KMV09}. For the case of exponential utility we obtain results that are, to the best of our knowledge, new. 
Even though the approach presented here is not (yet) mathematically rigorous, and, if made so, presumably equivalent to that of Kramkov and S{\^\i}rbu, it might still be interesting, as it sheds new light on the framework and is also straightforwardly applicable in a discrete time setting, which might be useful in practical applications.

In Section~\ref{sec:Interpretation}, we discuss our results, in particular the two problems mentioned above. The concept of implied drift that solves these is also introduced. As a nontrivial toy model that exemplifies the discussion, we use a jump diffusion with fixed jump size. We also compare the marginal utility price and hedge with those obtained in Merton's and the minimal variance approach. We conclude with a summary and an outlook.

\section{A heuristic derivation}
\label{sec:Derivation}

The basic idea of marginal utility based pricing and hedging is the following: Consider an investor with a concave utility function $U$, i.e., the expected utility for investment with time horizon $T$ is given by
\[
 u_t(x; \pi) = E[U(X^\pi_t)| X^\pi_t = x],
\]
where $X^\pi_t$ is the wealth process depending on some trading strategy $\pi$. This is maximized by the optimal investment strategy $\pi^*$:
\begin{equation*}
 u_t(x) = \sup_{\pi} u_t(x; \pi) = u_t(x; \pi^*).
\end{equation*}
Throughout this paper, we will consider trading strategies given by a space of functions $(t, x, s) \mapsto \pi_t(x,s)$ of time, wealth, and the asset price, that is equipped with some locally convex topology. Furthermore, we assume that the set of admissible trading strategies is an open subset, and that it contains a unique optimal investment strategy $\pi^*$.
This allows us to consider infinitesimal perturbations of $\pi^*$. Furthermore, $\pi^*$ should be such that the maximal expected utility is finite and such that the asset process is a local martingale under the measure $\Q$ defined by
\begin{equation}
\label{eq:Q}
\frac{\ud \Q}{\ud P} = \frac{U'_T(X_T^{\pi^*})}{E[U'_T(X_T^{\pi^*})]},
\end{equation}
where $P$ is the real-world measure. 
For conditions under which the last requirements are fulfilled, we refer to \cite{Sch01}.

If we now want to value a European claim with maturity $T$ and bounded payoff $C(S_T)$, where $S_t$ is the asset process, we force the investor to short an infinitesimal number $\varepsilon$ of them. For this she may charge a price $v^\varepsilon_t(x,s)$ per claim. It is the indifference price if
\begin{equation}
\label{eq:DefIndifference}
 u_t(x) = \sup_\pi E[U(X^\pi_T - \varepsilon C(S_T)) | X^\pi_t = x + \varepsilon v^\varepsilon_t(x,s), S_t = s].
\end{equation}
It means that the investor is willing to sell the options at a price $v^\varepsilon$, as this does not decrease her expected utility. The limit
\[
 v = \lim_{\varepsilon \to 0} v^\varepsilon
\]
is the \emph{marginal indifference price}. The marginal optimal hedge can be similarly characterized as the linear change of $\pi^*$ that is needed to achieve the maximum on the \rhs of \eqref{eq:DefIndifference}. Before we formalize this notion, we discuss the wealth process $X_t^\pi$ and the optimal investment strategy $\pi^*$ for the case of a jump diffusion and introduce functional differentiation, a technical tool we later employ.

\subsection{The wealth process and optimal investment}
\label{sec:OptimalInvestment}
The wealth process $X^\pi_t$ corresponding to the asset process \eqref{eq:AssetProcess}, given a trading strategy $\pi_t$, is given by
\begin{align*}
	\ud X^\pi_t & = \pi_t(X^\pi_{t_-}, S_{t_-}) \ud \log S_t \\
	& = \pi_t(X^\pi_{t_-}, S_{t_-}) \mu \ud t + \pi_t(X^\pi_{t_-}, S_{t_-}) \sigma \ud W_t + \pi_t(X^\pi_{t_-}, S_{t_-}) (e^{J_t} - 1) \ud N_t
\end{align*}
Here $\pi_t(x, s)$ denotes the wealth invested in the asset at time $t$, given that the total wealth is $x$ and the asset price is $s$. Note that no interest rate is present, so we are working in discounted units.

For a quantity $F_t(x, s)$, depending on time $t$, the wealth $x$ and the asset price $s$ that fulfills
\begin{equation}
\label{eq:piConsistency}
	F_t(x, s; \pi) = E[F_\tau (X^\pi_\tau, S_\tau; \pi) | X^\pi_t = x, S_t = s] \quad \forall t \leq \tau,
\end{equation}
one obtains the partial integro-differential equation (PIDE)
\begin{equation}
\label{eq:Fevolution}
\del_t F_t(x, s; \pi) + L^\pi F_t(x, s; \pi) = 0,
\end{equation}
where $L^\pi$ is the integro-differential operator defined by
\begin{align}
\label{eq:Lpi}
	L^\pi f_t(x, s) & = \mu \left\{ \pi_t(x, s) \del_x + s \del_s \right\} f_t(x, s)  \\
	& \quad + \tfrac{\sigma^2}{2} \left\{ {\pi_t(x, s)}^2 \del_x^2  + 2 \pi_t(x, s) s \del_x \del_s + s^2 \del_s^2 \right\} f_t(x, s) \nonumber\\
	& \quad + \int \left\{ f_t(x + \pi_t(x, s)(e^z-1), e^z s) - f_t(x, s) \right\} \ud \nu(z). \nonumber
\end{align}
Note that in \eqref{eq:Fevolution} we included the dependence on the trading strategy $\pi$. As $\pi$ is a function (of $t$, $x$ and $s$), $F$ is, apart from being a function of $t$, $x$ and $s$, also a \emph{functional}, i.e., a map from a space of functions to the real numbers. A useful tool for discussing extrema of such functionals are functional derivatives, which we briefly discuss in Section~\ref{sec:FunctionalDerivatives}.

In the absence of consumption, the expected utility $u_t(x; \pi)$ fulfills (\ref{eq:piConsistency}). For the maximal expected utility $u_t(x)$, the HJB equation
\begin{equation*}
\sup_\pi \left[ \del_t u_t(x) + L^\pi u_t(x) \right] = 0
\end{equation*}
holds, where the supremum is achieved by the optimal investment strategy $\pi^*$. Thus, the optimal investment strategy $\pi_t^*$ fulfills
\begin{equation*}
	\del_{\pi_t(x)} |_{\pi^*} L^\pi u_t(x) = 0.
\end{equation*}
Using the explicit form (\ref{eq:Lpi}) of $L^\pi$, we obtain
\begin{equation}
\label{eq:PiStar}
	\mu u'_t(x) + \pi^*_t(x) \sigma^2 u''_t(x) + \int u'_t(x^z) (e^z-1) \ud \nu(z) = 0,
\end{equation}
where we used the notation
\begin{equation}
\label{eq:Xz}
 x^z = x + \pi^*_t(x)(e^z-1)
\end{equation}
for the wealth after a jump. Having solved for $\pi^*$, we know that $u_t$ fulfills the PIDE
\begin{equation}
\label{eq:uPIDE}
	\del_t u_t(x) + L^{\pi^*} u_t(x) = 0.
\end{equation}
For investment with a time horizon $T$, the boundary condition is given by the utility $U$ at time $T$, i.e., $u_T(x) = U(x)$.

We now discuss the form of $\pi^*$ and $u_t$ in the two cases that will be of most interest to us, namely the case of constant relative or absolute risk aversion. Constant relative risk aversion is specified by a utility function
\begin{equation}
\label{eq:Upower}
 U(x) = x^{1-\beta} / (1-\beta), \quad \beta > 1.
\end{equation}
The limit $\beta \to 1$ corresponds to logarithmic utility, and the results below are also valid in that case. It can be shown that, up to an unimportant multiplicative constant, the expected utility $u_t$ is of the same form, i.e.,
\begin{equation*}
	u_t(x) = B_t x^{1-\beta} / (1-\beta).
\end{equation*}
Introducing the notation $\tilde \pi_t^*(x) = \pi_t^*(x) / x$, \eqref{eq:PiStar} becomes
\begin{equation}
\label{eq:TildePi}
	\mu -  \tilde \pi^*_t(x) \beta \sigma^2 + \int (e^z-1) (1 + \tilde \pi^*_t(x) (e^z-1))^{-\beta}  \ud \nu(z) = 0.
\end{equation}
We see that $\tilde \pi^*_t(x)$ is independent of $x$ and $t$. The optimal strategy is to invest a fixed fraction of the wealth in the asset. For precise conditions under which a solution to \eqref{eq:TildePi} exists, we refer to \cite{Nutz10b}.

We now discuss a special case which we will study explicitly in Section~\ref{sec:Interpretation} and in which \eqref{eq:TildePi} can be solved analytically:
\begin{example}
If only jumps of a certain size $J$ can happen, i.e., $\nu(z) = \lambda \delta(z-J)$, then \eqref{eq:TildePi} becomes
\begin{equation}
\label{eq:TildePiPower}
	\mu -  \tilde \pi^* \beta \sigma^2 + \lambda \tilde J (1 + \tilde \pi^* \tilde J)^{-\beta} = 0.
\end{equation}
Here we introduce the notation $\tilde J = e^J-1$ for the relative jump size. For logarithmic utility, i.e., for $\beta = 1$ this has an analytic solution:
\begin{equation}
\label{eq:TildePiLog}
	\tilde \pi^* = - \frac{1}{2 \tilde J} \left\{ 1 - \frac{\mu}{\sigma^2} \tilde J - \sqrt{\left( 1  - \frac{\mu}{\sigma^2} \tilde J \right)^2 + 4 \frac{\mu + \lambda \tilde J}{\sigma^2} \tilde J } \right\}.
\end{equation}
This has the expected behavior: $\tilde \pi^*$ always has the same sign as the average drift $\mu + \lambda \tilde J$. Also note that the expression under the square root is strictly positive, so that an optimal investment strategy
always exist for a fixed jump size. For $\beta > 1$, \eqref{eq:TildePiPower} can easily be solved numerically.
\end{example}

We briefly consider the case of constant absolute risk aversion, i.e., exponential utility 
\begin{equation}
\label{eq:Uexp}
 U(x)=-C e^{-\alpha x}, \quad \alpha > 0.
\end{equation}
As above, the expected utility is of the same form, where $C$ (but not $\alpha$) is time dependent. The optimal investment strategy $\pi^*_t(x)$ fulfills
 \begin{equation}
 \label{eq:PiExp}
 	\mu - \pi^*_t(x) \sigma^2 \alpha + \int (e^z-1) e^{- \alpha \pi^*_t(x) (e^z-1)} \ud \nu(z) = 0.
 \end{equation}
 A solution $\pi^*_t(x)$ of this equation will be independent of $x$ and $t$, so that the optimal strategy is to invest a fixed amount of wealth in the asset. Furthermore, it is antiproportional to the risk aversion $\alpha$, i.e., we can write
\begin{equation}
\label{eq:barPi}
 \pi^* = \bar \pi^* / \alpha
\end{equation}
for some constant $\bar \pi^*$. Noting that, for large $\beta$, $\tilde \pi_*(\beta)$ behaves as $\beta^{-1}$, one obtains, by inspection of \eqref{eq:TildePi} and \eqref{eq:PiExp} that
\begin{equation}
\label{eq:betaLimit}
 \lim_{\beta \to \infty} \beta \tilde \pi^*(\beta) = \bar \pi^*.
\end{equation}

Finally, we remark that by applying the PIDE \eqref{eq:uPIDE} to the terminal condition $W(x)=U'(x)$, one can show that for $\pi^*$ fulfilling \eqref{eq:PiStar}, one has $w_t = u'_t$, i.e.,
\begin{equation}
\label{eq:del_XU}
 u'_t(x, s) = E[U'(X^{\pi^*}_T) | X^{\pi^*}_t = x, S_t = s].
\end{equation}

\subsection{Functional derivatives}
\label{sec:FunctionalDerivatives}
We briefly introduce the concept of functional derivatives as a special case of directional derivatives, cf. \cite{Hamilton, lcs}. Let $F$ be a functional, i.e., a continuous map $U \to \R$, where $U$ is an open subset of a space $X$ of functions\footnote{In order to define these notions, $X$ has to be equipped with a topology, which we assume to be locally convex.}. Then $F$ is called \emph{differentiable} at $f \in U$ in the direction $h \in X$ if the limit
\[
 \skal{\delta F(f)}{h} := \lim_{t \to 0} \frac{F(f+th) - F(f)}{t}
\]
exists. It is called \emph{continuously differentiable} (or $C^1$) on $U$ if the limit exists for all $f \in U$, $h \in X$ and if $\delta F: U \times X \to \R$ is continuous. If $F$ is $C^1$, then $\delta F(f): X \to \R$ is linear. Many of the usual theorems of differential calculus hold, in particular the fundamental theorem. It follows that a necessary condition for a $C^1$ functional to have a local maximum in $f$ is the vanishing of $\delta F(f)$.

The second derivative can be defined as the derivative of the first derivative, \wrt $f$, i.e.,
\[
 \skal{\delta^2 F(f)}{h \otimes k} := \lim_{t \to 0} \frac{\skal{\delta F(f+tk)}{h} - \skal{\delta F(f)}{h}}{t}.
\]
We say that $F$ is $C^2$ on $U$ if the limit exists for all $f \in U$ and $h,k \in X$ and is a continuous map $\delta^2 F: U \times X \times X \to \R$. In that case $\delta^2 F(f)$ is bilinear and symmetric. This generalizes to derivatives of arbitrary order. There is a Taylor formula from which it follows that a sufficient condition for a $C^2$ function to have a local maximum in $f \in U$ is that $\delta F(f) = 0$ and $\skal{\delta^2 F(f)}{h \otimes h} < 0$ for all $h \in X$.

If $F$ is $C^1$, then $\delta F(f)$ is a continuous linear functional on $X$, so $\delta F(f)$ is a distribution. Similarly, if $F$ is $C^2$, then $\delta^2 F(f)$ is a symmetric bi-distribution. Employing the familiar abuse of notation to express a distribution in terms of an integral kernel, we sometimes write
\[
 \skal{\delta F(f)}{h} = \int \delta_{f(x)} F(f) h(x) \ud x,
\]
and analogously for the higher order derivatives.


The functionals that we want to differentiate below are solutions to a PIDE of the form \eqref{eq:Fevolution}, which we want to differentiate \wrt $\pi$. More precisely, let $w$ be a solution to the PIDE
\[
 \del_t w_t(x, s; \pi) + M^\pi w_t(x, s; \pi) = 0,
\]
where $M^\pi_t$ is an integro-differential operator that depends on $\pi$. We will want to compute
\[
 \delta_{\pi_t(x,s)} w_t(x, s; \pi),
\]
i.e, compute the change in $w_t(x,s; \pi)$ if $\pi$ is perturbed at the same point, namely at time $t$, wealth $x$ and asset price $s$. We assume that the PIDE is solved backwards in time from some terminal condition. Formally, we thus have
\[
 w_t(x, s; \pi) = w_{t+\ud t}(x, s; \pi) +  M^\pi w_t(x, s; \pi) \ud t.
\]
The effect of turning on a perturbation $\pi'$ of $\pi$ that is localized around $x, s$ and in the time interval $[t, \ud t)$, can thus be computed by differentiating $M^\pi$ \wrt $\pi$. One thus obtains
\[
  \skal{\delta_\pi w_t(x, s; \pi)}{\pi'} = \del_\pi M^\pi w_t(x, s; \pi) \pi'(x, s) \ud t.
\]
Here $\del_\pi$ only acts on the operator $M^\pi$. The limit where $\pi'$ tends to a Dirac $\delta$ in time corresponds to $\ud t \to 0$, $\pi' \sim \ud t^{-1}$. Hence, we obtain
\begin{equation}
\label{eq:FunctionalDiff}
 \delta_{\pi_t(x, s)} w_t(x, s; \pi) = \del_{\pi_t(x,s)} M^\pi w_t(x, s; \pi).
\end{equation}

In the following, we assume that the expected utility is $C^2$ in a neighborhood $U$ of $\pi^*$. This is the case if
\[
 \int u''_t(x+\pi_t(e^z-1)) (e^z-1)^2 \ud \nu(z) < \infty
\]
for $\pi \in U$. For $\pi = 0$, this means that the jump distribution must have a finite second moment.

\subsection{The marginal indifference price}
We want to determine the marginal indifference price $v$ from \eqref{eq:DefIndifference}. As the perturbation is infinitesimally small, we can assume that the trading strategy $\pi^\varepsilon$ that achieves the maximum on the \rhs of \eqref{eq:DefIndifference} fulfills $\pi^\varepsilon = \pi^* + \varepsilon \bar \pi + \order(\epsilon^2)$. We also have $v^\varepsilon = v + \order(\varepsilon)$. Thus, expanding \eqref{eq:DefIndifference} in $\varepsilon$, we obtain
\begin{multline*}
 u_t(x) = u_t(x) + \varepsilon v_t(x,s) \del_x u_t(x,s) - \varepsilon E[ U'(X^{\pi^*}_T) C(S_T) | X^{\pi^*}_t = x, S_t = s] \\
 + \varepsilon \skal{\delta_\pi u_t(x; \pi^*)}{\bar \pi} + \order(\epsilon^2),
\end{multline*}
To obtain the fourth term on the r.h.s., we used functional differentiation \wrt $\pi$.
This term vanishes, since $\pi^*$ is optimal. Equating the remaining terms of first order in $\epsilon$, one obtains \cite{Davis}
\begin{equation}
\label{eq:MarginalPrice}
 v_t(x, s) = \frac{E[ U'(X^{\pi^*}_T) C(S_T) | X^{\pi^*}_t = x, S_t = s]}{u'_t(x) }.
\end{equation}
It follows that in order to determine the marginal indifference price $v$ it suffices to know $\pi^*$, i.e., one does not have to solve the full optimization problem. Using the tower property, \eqref{eq:MarginalPrice} can be expressed as an expected value for quantities at times $\tau$ with $t < \tau \leq T$:
\begin{equation}
\label{eq:MarginalPriceTau}
	v_t(x, s) = \frac{E[ u'_\tau(X^{\pi^*}_\tau) v_\tau(X^{\pi^*}_\tau, S_\tau) | X^{\pi^*}_t = x, S_t = s]}{u'_t(x) }.
\end{equation}
In this form, the pricing problem can be solved backwards in time with the payoff as the terminal condition. In continuous time, the limit $\tau = t + \ud t$ will yield a partial (integro-) differential equation.

In the case of jump diffusion, one obtains from \eqref{eq:MarginalPriceTau} the PIDE
\begin{equation}
\label{eq:PricingPIDE}
	\del_t v_t(x, s) + L^\Q v_t(x, s) = 0,
\end{equation}
where $L^\Q_t$ is the integro-differential operator defined by
\begin{align}
\label{eq:LQ}
	L^\Q f_t(x, s) & = \mu^\Q \left\{ \pi^*_t(x, s) \del_x + s \del_s \right\} f_t(x, s)  \\
	& \quad + \tfrac{\sigma^2}{2} \left\{ {\pi^*_t(x, s)}^2 \del_x^2  + 2 \pi^*_t(x, s) s \del_x \del_s + s^2 \del_s^2 \right\} f_t(x, s) \nonumber\\
	& \quad + \int \left\{ f_t(x^z, e^z s) - f_t(x, s) \right\} \ud \nu^\Q(x; z). \nonumber
\end{align}
where we used the notation \eqref{eq:Xz} and changed the drift and the jump distribution \wrt $L^\pi$, cf. \eqref{eq:Lpi}, as
\begin{align}
\label{eq:nuQ}
 \ud \nu^\Q(x; z) & = \frac{u'_t(x^z)}{u'_t(x)} \ud \nu(z), \\
\label{eq:muQ}
 \mu^\Q(x) & = - \int (e^z-1) \ud \nu^\Q(x; z).
\end{align}
Here the adjusted drift and jump distribution define the risk-neutral measure $\Q$, cf. \eqref{eq:Q}.

Let us briefly discuss the intuition behind \eqref{eq:nuQ}. Recall that $U$, and thus also $u$, is concave, i.e., $u'(y) < u'(x)$ for $y > x$. Suppose the asset has, on average, positive returns. Then the optimal investment strategy will be to invest in the asset, i.e., $\pi^* > 0$. Then, for a downward jump, $z<0$, we have $x^z < x$. It follows that the fraction on the \rhs of \eqref{eq:nuQ} is greater than one, so that downward jumps become more (and upward jumps less) frequent. The opposite happens for $\pi^* < 0$. The economic rationale behind this is the following: If the investor is invested in the asset, she is exposed to the risk of downward jumps. She will thus seek remuneration for taking even more downward jump risk. On the other hand, she is also exposed to the risk of no upward jumps happening. She is thus willing to sell a claim that is exposed to upward jump risk with a discount. Finally \eqref{eq:muQ} serves to set the average drift to zero.

\begin{example}
If the process is a pure diffusion, i.e., $\lambda = 0$, the terms in the first and the third line in \eqref{eq:LQ} vanish. All other terms that have some $x$-dependence involve $\del_x$. Thus, if the terminal condition is independent of $x$, as for a payoff, the marginal indifference price is also independent of $x$ and one recovers the Black--Scholes PDE, in discounted units.
\end{example}

From our discussion in Section \ref{sec:OptimalInvestment}, we know that in the case of constant relative or absolute risk aversion $u'_t(x^z) / u'_t(x)$ is independent of $x$. Thus, for these types of utility, the only terms in $L^\Q$ that depend on $x$ are those that involve at least one $\del_x$. It follows that if the terminal condition is independent of $x$, the solution to \eqref{eq:PricingPIDE} will also be independent of $x$. Since by definition the payoff only depends on $s$, the marginal indifference price is independent of $x$ for constant relative or absolute risk aversion \cite{KS06}.
We thus obtain
\begin{proposition}
 In the case of power utility, \eqref{eq:Upower}, the marginal indifference price is a solution to the PIDE
\begin{multline}
\label{eq:PricingPIDEPower}
	0 = \del_t v_t(s) + \left\{ \int \{ e^z - 1\} \left(1+\tilde \pi^*(e^z-1) \right)^{-\beta} \ud \nu(z) \right\} s \del_s v_t(s) + \tfrac{\sigma^2}{2} s^2 \del_s^2 v_t(s) \\
	+ \int \left\{ v_t(e^z s) - v_t(s) \right\} \left(1+\tilde \pi^*(e^z-1) \right)^{-\beta} \ud \nu(z),
\end{multline}
where $\tilde \pi^*$ is a solution to \eqref{eq:TildePi}. In the case of exponential utility, \eqref{eq:Uexp}, the marginal indifference price is a solution to the PIDE
\begin{multline}
\label{eq:PricingPIDEExp}
	0 = \del_t v_t(S) + \left\{ \int \{ e^z - 1\} e^{- \bar \pi^* (e^z-1)} \ud \nu(z) \right\} s \del_s v_t(s) + \tfrac{\sigma^2}{2} s^2 \del_s^2 v_t(s) \\
	+ \int \left\{ v_t(e^z s) - v_t(s) \right\} e^{- \bar \pi^* (e^z-1)} \ud \nu(z),
\end{multline}
where $\bar \pi^*$ is given by \eqref{eq:barPi}.
\end{proposition}

\begin{remark}
A special case of \eqref{eq:PricingPIDEPower} was found in \cite{Lewis}. There, it is assumed that all market participants have power utility, and so the market-clearing utility must also be of this form. Furthermore, the market is invested fully in the asset, the positions in cash and options cancel each other\footnote{This implies that the model is only applicable to an ``index'' that comprises the whole market, i.e., in principle equities, commodities, real estate, etc.}. This corresponds to $\tilde \pi^*(x) = 1$ in the present setting, which, inserted in \eqref{eq:PricingPIDEPower}, gives the PIDE of \cite{Lewis}.
\end{remark}

\subsection{The marginal optimal hedge}

We now want to study the marginal optimal hedge corresponding to the marginal indifference price. In the previous section, we expressed the trading strategy that maximizes the \rhs of \eqref{eq:DefIndifference} as $\pi^\varepsilon = \pi^* + \varepsilon \bar \pi + \order(\varepsilon^2)$. We define the marginal optimal hedge $\hat \pi$ as
\[
 \hat \pi = \bar \pi + v \del_x \pi^*.
\]
The idea behind this definition is the following: We want to determine the change in the optimal trading strategy that is caused by the option trade. Thus, we wish the investor to invest optimally as she would do without the trade plus some correction which we wish determine. The purpose of the second term on the \rhs of the above equation is to cancel the shift in the optimal investment strategy that is caused by the payment of the option price $v$.

As $\pi^\varepsilon$ is the optimizer on the \rhs of \eqref{eq:DefIndifference}, the functional derivative at this point should vanish:
\begin{equation*}
 \delta_{\pi_{t'}(x',s')} E[U(X^{\pi^\varepsilon}_T - \varepsilon C(S_T)) | X^{\pi^\varepsilon}_t = x + \varepsilon v^\varepsilon_t(x,s), S_t = s] = 0 \quad \forall (t', x', s').
\end{equation*}
Expanding this in $\varepsilon$, one obtains
\begin{align}
\label{eq:expansion}
 0 & = - \delta_{\pi_{t'}(x',s')} E[U'(X^{\pi^*}_T) C(S_T) | X^{\pi^*}_t = x, S_t = s] \\
   & \quad + v_t(x,s) \delta_{\pi_{t'}(x',s')} \del_x u_t(x; \pi^*) \nonumber \\ 
   & \quad + \int \bar \pi_{t''}(x'', s'') \delta_{\pi_{t'}(x',s')} \delta_{\pi_{t''}(x'',s'')} u_t(x; \pi^*) \ud t'' \ud x'' \ud s''. \nonumber
\end{align}
The two derivatives in the second line commute, so by the optimality of $\pi^*$, this term vanishes.

We now want to argue that the second order functional derivative in the third term vanishes unless $t' = t''$. Here the optimality (and the implicitly assumed Markov property of $S$) is crucial. Assume that $t'' > t'$. Then we may find $\tau$ such that $t' < \tau < t''$. We have, by the tower property
\[
  u_t(x; \pi) = E[U(X^{\pi}_T) | X^{\pi}_t = x] =  E[ E[U(\tilde X^{\pi}_T) | \tilde X^{\pi}_\tau = X^\pi_\tau] | X^{\pi}_t = x].
\]
Here we introduced the notation $\tilde X$ for the process from $\tau$ to $T$ in order to distinguish in from the process $X$ on which it is conditioned at $\tau$. Changing $\pi$ at $t'$ only affects the process $X$, while changing $\pi$ at $t''$ only affects $\tilde X$. In particular, the derivative \wrt $\pi_{t''}(x'', s'')$ can be pulled inside the outer expected value and the derivative \wrt $\pi_{t'}(x', s')$ does not act on the inner expected value. But the derivative \wrt $\pi_{t''}(x'', s'')$ of the inner expected value, evaluated at $\pi^*$, vanishes, by optimality. Thus, the second order functional derivative in the third term on the \rhs of \eqref{eq:expansion} vanishes unless $t' = t''$. 

For well behaved processes (in particular there should be no predetermined jump times), the second order functional derivative at equal times $t' = t''$ vanishes unless $(x', s') = (x'', s'')$. The intuitive reason is that a change of the trading strategy at $(t', x', s')$ can only affect paths that are at $(x', s')$ at time $t'$. But a path can not be at $(x',s')$ and $(x'', s'')$ at the same time unless $(x', s') = (x'', s'')$. We may thus write
\begin{multline*}
 \delta_{\pi_{t'}(x',s')} \delta_{\pi_{t''}(x'',s'')} u_t(x; \pi^*) \\ = \delta_D(t' - t'') \delta_D(x' - x'') \delta_D(s' - s'') \delta_{\pi_{t'}(x',s')}^2 u_t(x; \pi^*).
\end{multline*}
Here $\delta_D$ denotes the Dirac $\delta$ distribution and the second order functional derivative on the \rhs is implicitly defined by this equation.

Applying this to \eqref{eq:expansion}, we obtain
\begin{multline*}
 \delta_{\pi_{t'}(x',s')} E[U'(X^{\pi^*}_T) C(S_T) | X^{\pi^*}_t = x, S_t = s] \\
 = \left( \hat \pi_{t'}(x', s') - v_{t'}(x', s') \del_x \pi^*_{t'}(x') \right) \delta_{\pi_{t'}(x',s')}^2 u_t(x; \pi^*).
\end{multline*}
For a constant payoff $P$, $v = P$ and $\hat \pi$ vanishes, so that we have
\begin{equation*}
 \del_x \pi^*_{t'}(x') \delta_{\pi_{t'}(x',s')}^2 u_t(x; \pi^*) = - \delta_{\pi_{t'}(x',s')} E[U'(X^{\pi^*}_T) | X^{\pi^*}_t = x, S_t = s].
\end{equation*}
This is valid for all $(t', x', s')$. In particular, we may choose $(t', x', s') = (t, x, s)$, and with \eqref{eq:MarginalPrice} and \eqref{eq:del_XU} we obtain
\begin{equation}
\label{eq:Proposition_2}
 \hat \pi_t(x, s) = \frac{u'_t(x, s) \delta_{\pi_{t}(x,s)} \frac{E[U'(X^{\pi^*}_T) C(S_T) | X^{\pi^*}_t = x, S_t = s]}{E[U'(X^{\pi^*}_T) | X^{\pi^*}_t = x, S_t = s]} }{\delta_{\pi_{t}(x,s)}^2 u_t(x; \pi^*)}.
\end{equation}
Note that the expression that is functionally differentiated in the numerator is the marginal indifference price. Thus, this equation has a straightforward economic interpretation: The functional derivative in the numerator gives the marginal gain in the price $v$ one can generate by shifting $\pi$ from the optimal trading strategy $\pi^*$. The factor in front of the functional derivative converts this into a marginal gain in utility. This gain in utility stemming from $v$ is to be balanced by the loss in utility that is incurred to the wealth process (without the claim) by deviating from $\pi^*$, which one finds in the denominator\footnote{One might think of the following analogy: Let $f$ be a function with a local maximum at $x^*$ and $f''(x^*) < 0$. Perturbing $f(x)$ by subtracting $\epsilon g(x)$, the new maximum is found at $x_\epsilon^* = x^* + \epsilon g'(x^*)/f''(x^*) + \order(\epsilon^2)$.}.

We also note the similarity of this formula with the sensitivities that are used in $\Delta$-hedging. However, in the present case, one does not differentiate the option value \wrt the asset price but \wrt the trading strategy and weights the result with derivatives of expected utility. This representation of the marginal optimal hedge might be useful in numerical calculations, where it could be used to compute $\hat \pi$, and simultaneously $v$, backwards in time in a discrete-time setting.

We may now evaluate \eqref{eq:Proposition_2} for the case of a jump diffusion. We note that the two quantities that are functionally differentiated are solutions to PIDEs that depend on $\pi$. We may thus proceed as discussed in Section~\ref{sec:FunctionalDerivatives}, i.e., we apply \eqref{eq:FunctionalDiff}. As the derivation of \eqref{eq:PricingPIDE} from \eqref{eq:MarginalPrice} did not make use of the optimality of $\pi^*$, the functional derivative in the numerator may be computed by differentiating $L^\Q$ \wrt $\pi$. Similarly, for the computation of the functional derivative in the denominator, we twice differentiate $L^\pi$ \wrt $\pi$. Restricting again to the case of constant relative or absolute risk aversion, one obtains\footnote{The hedge (and also the price) for power utility is implicitly contained in \cite{KMV09}.}
\begin{proposition}
In the case of power utility, \eqref{eq:Upower}, the marginal optimal hedging strategy is given by
\begin{equation}
\label{eq:OptimalMarginalHedgePower}
	\hat \pi_t(s) = s \frac{ \sigma^2 \del_s v_t(s) + \int \frac{ v_t(e^z s) - v_t(s)}{(e^z-1)s} \left( e^z - 1 \right)^2 \left( 1 + \tilde \pi^*(e^z-1) \right)^{-\beta-1} \ud \nu(z)}{\sigma^2 + \int \left( e^z - 1 \right)^2 \left( 1 + \tilde \pi^*(e^z-1) \right)^{-\beta-1} \ud \nu(z)},
\end{equation}
where $\tilde \pi^*$ is a solution to \eqref{eq:TildePi}. In the case of exponential utility, \eqref{eq:Uexp}, the marginal optimal hedge is
\begin{equation}
\label{eq:OptimalMarginalHedgeExp}
	\hat \pi_t(s) = s \frac{ \sigma^2 \del_s v_t(s) + \int \frac{ v_t(e^z s) - v_t(s)}{(e^z-1)s} \left( e^z - 1 \right)^2 e^{-\bar \pi^*(e^z-1)} \ud \nu(z)}{\sigma^2 + \int \left( e^z - 1 \right)^2 e^{-\bar \pi^*(e^z-1)} \ud \nu(z)},
\end{equation}
where $\bar \pi^*$ is given by \eqref{eq:barPi}.
\end{proposition}


As the expected utility is $C^2$ by assumption, cf. Section~\ref{sec:FunctionalDerivatives}, the integral in the denominator in \eqref{eq:OptimalMarginalHedgePower} and \eqref{eq:OptimalMarginalHedgeExp} is finite. It follows that also the integrals in the numerator are finite, by the boundedness of $v$.

\begin{example}
In the pure diffusion case $\lambda = 0$ one recovers Black--Scholes $\Delta$ hedging, $\hat \pi = s \del_s v$. But if a jump component is present, the marginal optimal hedge is not given by $s \del_s v$. Instead, it optimally balances diffusion and jump risk, given the specified utility function.
\end{example}

\begin{remark}
\label{rem:RiskPremium}
Taking the marginal optimal hedge \eqref{eq:OptimalMarginalHedgePower}, \eqref{eq:OptimalMarginalHedgeExp} as starting point and following the derivation of \eqref{eq:MertonPIDE} from \eqref{eq:Portfolio}, one does in general not recover the PIDE \eqref{eq:PricingPIDEPower}, \eqref{eq:PricingPIDEExp} for the marginal indifference price. This is not surprising, since in the present framework the investor wants to be compensated for taking risk.
\end{remark}

\subsection{Minimal variance pricing and hedging}
The basic idea of minimal variance pricing and hedging was briefly discussed in the introduction. Here, we content ourselves with giving the corresponding price and hedge for our jump diffusion process. For the minimal variance price, one finds the following PIDE \cite{ColwellElliott93}:
\begin{multline}
\label{eq:PricingPIDE_MV}
	\del_t v_t(s) - \left\{ \int \left\{ e^z - 1 \right\} \left\{ 1-\alpha(e^z-1) \right\} \ud \nu(z) \right\} s \del_s v_t(s) + \tfrac{\sigma^2}{2} s^2 \del_s^2 v_t(s) \\ + \int \left\{ v_t(e^z s) - v_t(s) \right\} \left\{ 1-\alpha(e^z-1) \right\} \ud \nu(z) = 0.
\end{multline}
Here $\alpha$ is a generalization of the market price of risk and is given by
\begin{equation*}
  \alpha = \frac{\mu + \int (e^z -1) \ud \nu(z)}{\sigma^2 + \int (e^z -1)^2 \ud \nu(z)}.
\end{equation*}
Note that the new jump measure in \eqref{eq:PricingPIDE_MV} gives negative frequencies for jumps with $\alpha (e^z-1) > 1$. For $\alpha > 0$, this condition will always be fulfilled for unbounded upward jump distributions, which includes Merton's log-normal jump distribution \cite{Merton}. 

For the corresponding hedge one obtains
\begin{equation}
\label{eq:mvHedge}
	\theta_t(s) = \frac{\sigma^2 \del_s v_t(s) + \int \frac{v_t(e^z s)-v_t(s)}{(e^z-1)s} (e^z-1)^2 \ud \nu(z)}{\sigma^2 + \int (e^z-1)^2 \ud \nu(z)}.
\end{equation}

\begin{remark}
\label{rem:RiskCompensation}
Taking $\theta$ as hedging strategy and following the derivation of Merton's formula \eqref{eq:MertonPIDE} from \eqref{eq:Portfolio}, one recovers the pricing PIDE \eqref{eq:PricingPIDE_MV}. This shows that with minimal variance hedging one tries to minimize risk (as measured by the variance), but one is not compensated for it. This is in contrast to the setting of utility maximization, cf. Remark \ref{rem:RiskPremium}. Possible modifications of the framework to include also a risk premium are discussed, e.g., in~\cite{WilmottAhn}.
\end{remark}

\subsection{The approach of Kramkov and S{\^\i}rbu}

We want to briefly compare the (as yet heuristic) framework of functional differentiation presented here with the rigorous approach of Kramkov and S{\^\i}rbu \cite{KS06, KS07}. There, one restricts to utility functions with bounded relative risk aversion, of which power utility is a special case. Optimal hedging strategies are defined by their wealth process. The optimal wealth process for an initial capital $x$ and a quantity $q$ of claims is denoted by $X(x,q)$. The utility-based wealth process $G(x,q)$ is defined as
\[
 G(x,q) = X(c(x,q), 0) - X(x,q),
\]
where $c$ is the indifference price. The marginal optimal hedge $H$ is defined as the derivative of $G$ \wrt $q$ at $q=0$. Hence, the definition of the marginal optimal hedge presented here is based on the same idea as the definition of Kramkov and S{\^\i}rbu. 

Let us see whether the two notions coincide in the case of power utility. By \cite[Thm. 1]{KS07}, the process $H$ is given by (for initial capital $x=1$)
\[
 H_t = X^{\pi^*}_t \left( V_0 + M \right),
\]
where $V_0$ is the marginal indifference price and the process $M$ is the minimizer of the optimization problem
\[
 c = \inf_{M} E_R \left[ \frac{U''(X_T^{\pi^*})}{X_T^{\pi^*} U'(X_T^{\pi^*})}  \left( X_T^{\pi^*} (V_0 + M_T) - C(S_T) \right)^2 \right].
\]
Here $R$ is the measure given by
\[
 \frac{\ud R}{\ud P} = X_T^{\pi^*} \frac{U'(X_T^{\pi^*})}{u'}.
\]
Following \cite{KMV09}, one may write this as a minimal variance hedging problem under the measure $\tilde R$ defined by
\[
 \frac{\ud \tilde R}{\ud R} = \frac{U''(X_T^{\pi^*})}{X_T^{\pi^*} U'(X_T^{\pi^*})} E_R\left[ \frac{U''(X_T^{\pi^*})}{X_T^{\pi^*} U'(X_T^{\pi^*})} \right]^{-1},
\]
provided that the \rhs is a uniformly integrable martingale. As shown in \cite{KMV09}, this is the case for power utility if the jump distribution has a finite second moment, a condition that we had to impose, too, cf.\ the discussion at the end of Section~\ref{sec:FunctionalDerivatives}. Then one can use the minimal variance hedging formula \eqref{eq:mvHedge} with this measure. One indeed obtains \eqref{eq:OptimalMarginalHedgePower}.

%


\section{Discussion}
\label{sec:Interpretation}

We will now discuss the results obtained so far. In order to exemplify the findings, we use the toy model of a jump diffusion with a fixed jump size $J$. This model is analytically tractable \cite{Merton}. Re-introducing the risk-free rate $r$, we have to solve a PIDE of the form
\begin{equation*}
	\del_t v_t(s) + \tfrac{\sigma^2}{2} s^2 \del_s^2 v_t(s) + \{ r - \bar \lambda \tilde J \} s \del_s v_t(s) - r v_t(s) 
	+ \bar \lambda \left\{ v_t(e^J s) - v_t(s) \right\} = 0,
\end{equation*}
where we used $\tilde J = e^J -1$. The only difference between the different methods and utilities lies in the value of $\bar \lambda$ that is employed. The above is solved by
\begin{equation*}
	v_t(s) = \sum_{k = 0}^\infty \frac{( \bar \lambda (T-t))^k e^{- \bar \lambda (T-t)}}{k!} v_t(e^{k J} s, r, \bar \lambda \tilde J)
\end{equation*}
where $v_t(s, r, q)$ is the Black--Scholes price for the claim, given a risk-free rate $r$ and a dividend yield $q$.

\subsection{The price}
\begin{figure}
	\centering
		\includegraphics[scale = 0.5]{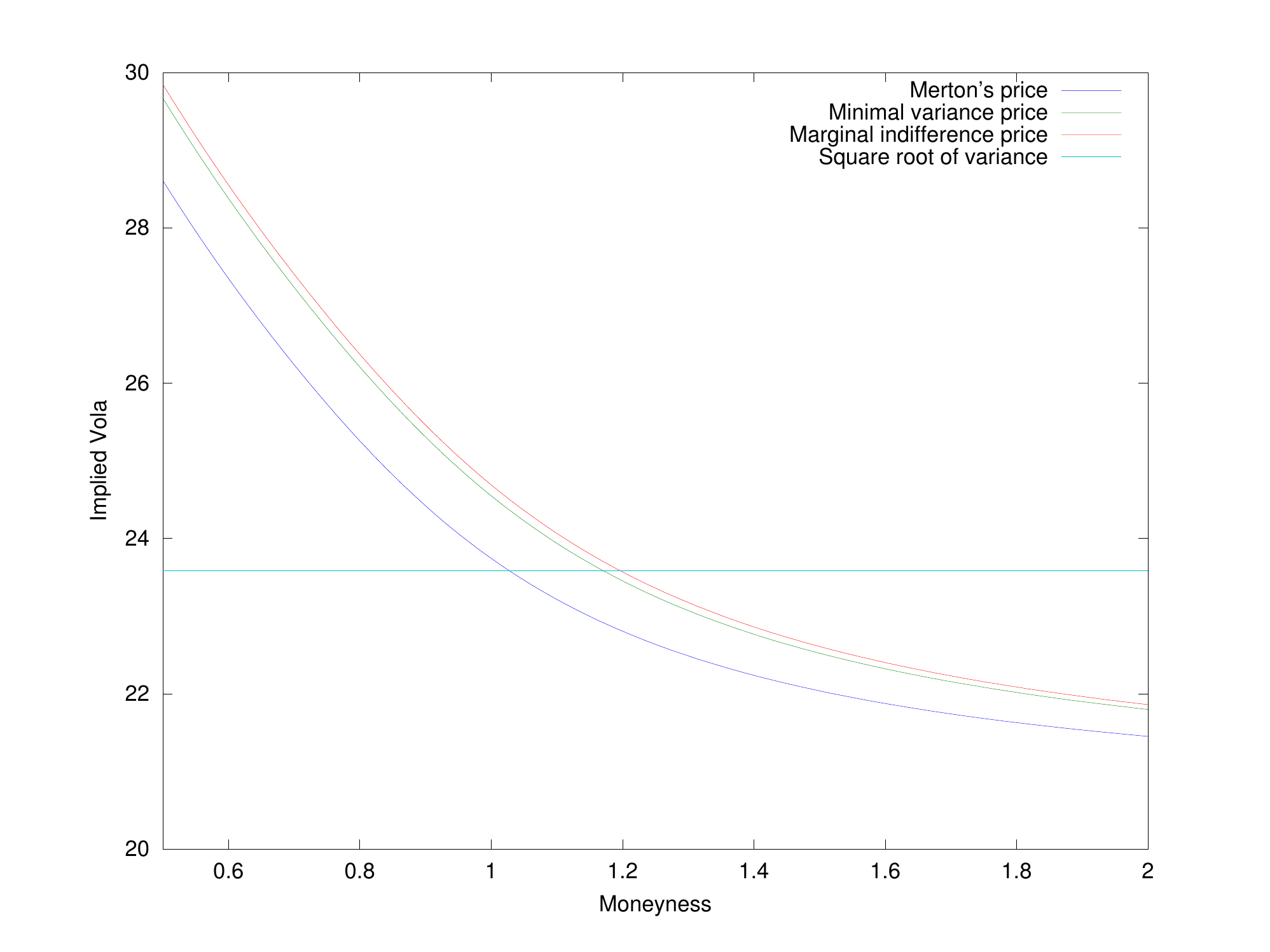}
	\caption{Implied volatilities for a jump diffusion with fixed jump size $\tilde J= -0.25$ using the different pricing methods for the parameters $\sigma = 0.2$, $r=0$, $\lambda = 0.25$, $\tilde \mu = \mu + \lambda  \tilde J = 0.05$ and $T = 1$. For the marginal indifference price logarithmic utility was used.}
		\label{fig:Price_PosRet}
\end{figure}

In order to get a feeling for the magnitude of the effect, we compare the marginal indifference price for logarithmic utility with Merton's and the minimal variance price. We use a process with $\lambda = 0.25$, $\tilde J = - 0.25$, i.e., on average there is a jump of $-25 \%$ every four years. For the marginal indifference price, the relevant value for $\bar \lambda$ is obtained from \eqref{eq:TildePiLog} and \eqref{eq:nuQ}. The price for a put (converted to implied volatilities) is shown in Figure~\ref{fig:Price_PosRet}. As a reference, the square root of the annualized variance is indicated. We see that the marginal indifference price and the minimal variance price are quite close together, but the difference to Merton's price is notable. For a moneyness of 0.5, it corresponds to a price difference of 40\%. 

\begin{figure}
	\centering
		\includegraphics[scale = 0.5]{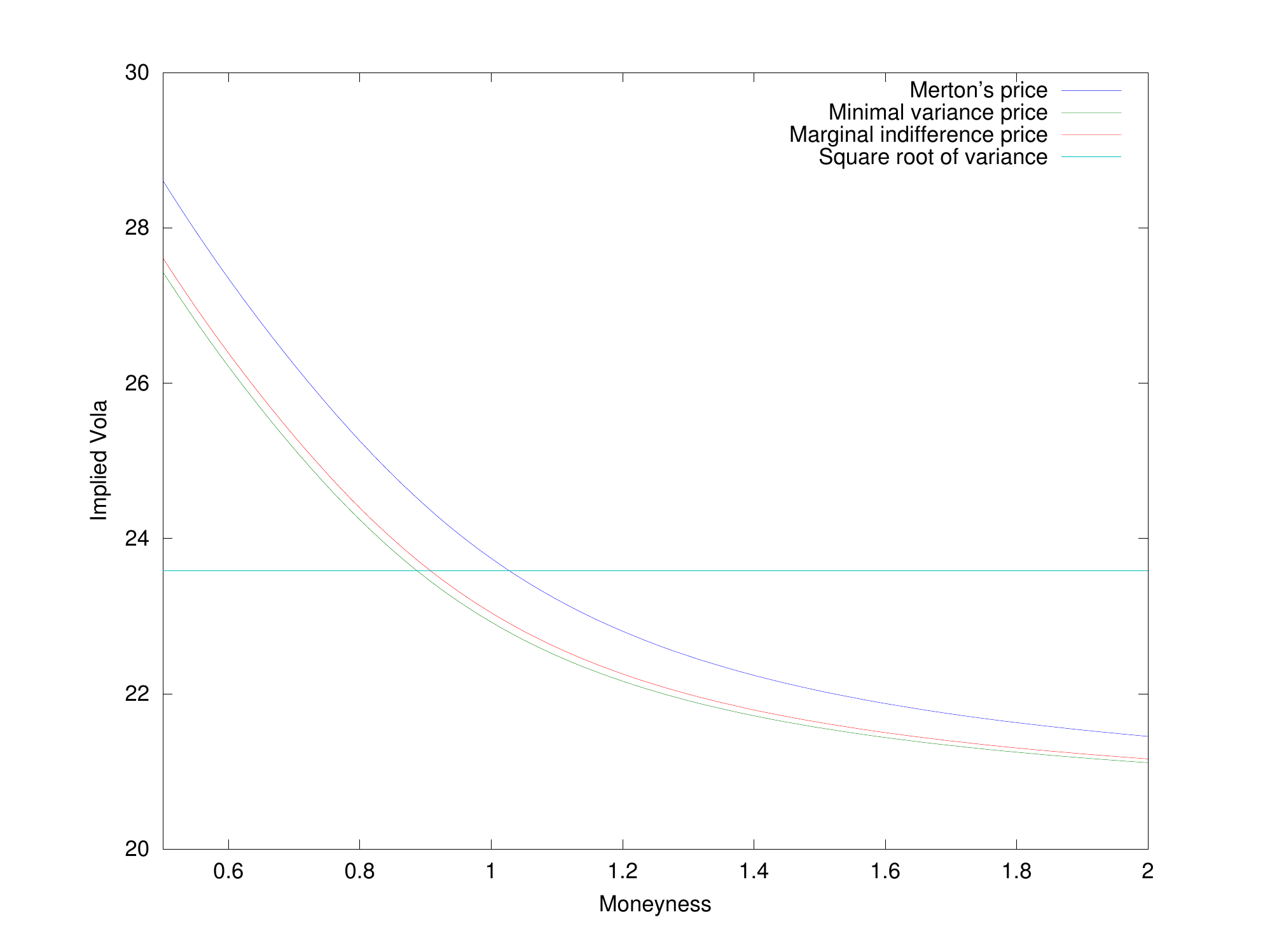}
	\caption{Same as Figure \ref{fig:Price_PosRet}, but with negative expected returns, $\tilde \mu = \mu + \lambda  \tilde J = -0.05$.}
		\label{fig:Price_NegRet}
\end{figure}

That the marginal indifference price and the minimal variance price are above Merton's price is not a generic feature, but depends on the average drift of the asset. This is illustrated in Figure \ref{fig:Price_NegRet}, which shows the same plot as before, but with an expected drift $\tilde \mu = - 0.05$. Now the marginal indifference price and the minimal variance price are below Merton's price. This can be understood as follows: If the expected drift is positive, the investor will be invested in the asset. Since jumps are always downwards in our model, she is exposed to jump risk. Writing a put on the asset in this situation enlarges this exposure. She will thus ask for a risk premium. On the other hand, if the expected drift is negative, the investor is short the asset and is then exposed to the risk of no jumps happening. Writing a put in this situation diminishes the exposure to this risk. Thus, she can sell the put with a discount. This strong dependence on the drift seems to limit the practical applicability of the framework, as it is very hard to estimate.

\begin{figure}
	\centering
		\includegraphics[scale = 0.5]{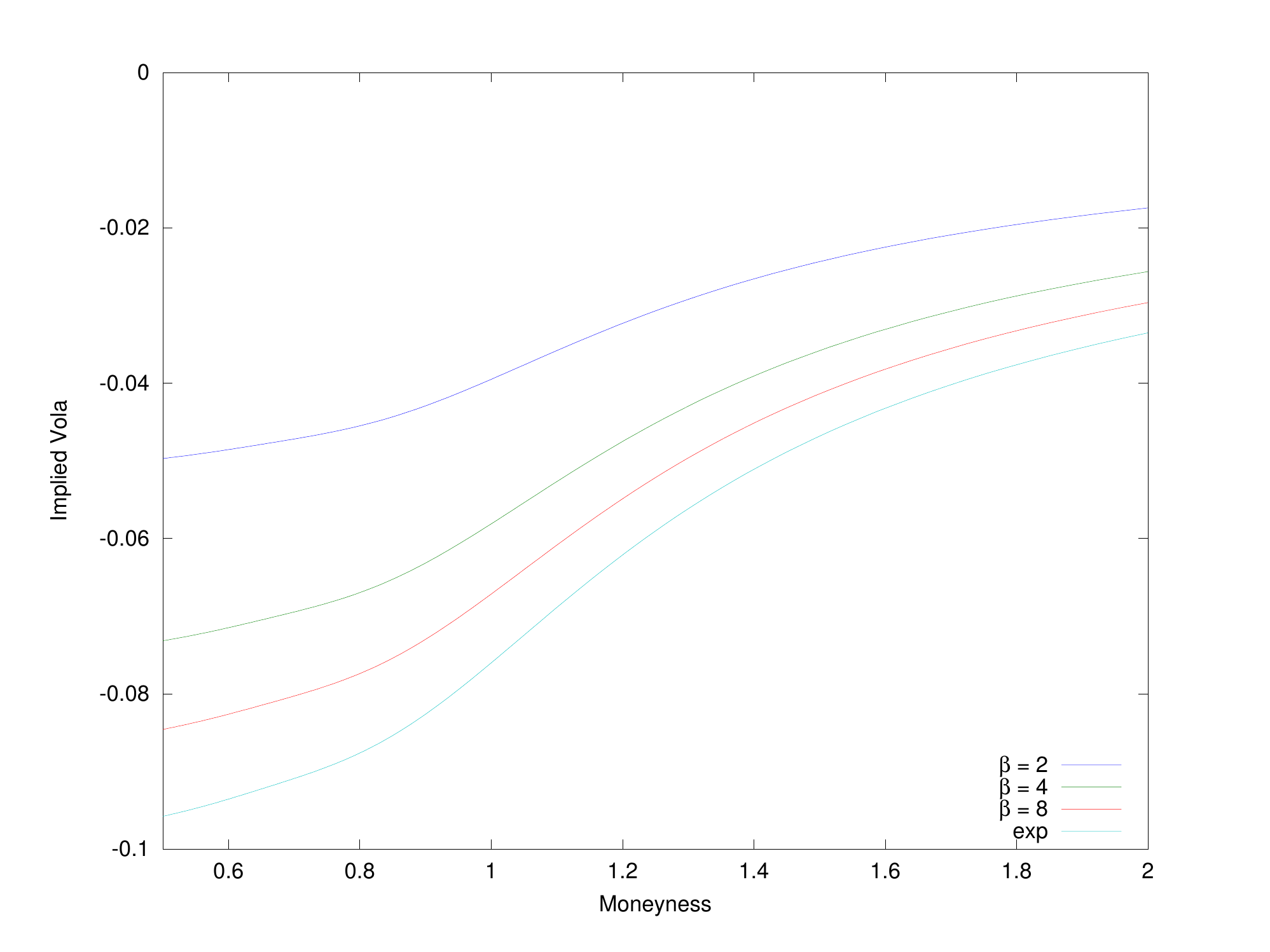}
	\caption{Differences of implied volatilities of the marginal indifference prices for different risk aversions and the one obtained for logarithmic utility ($\beta = 1$) for the same parameters as in Figure~\ref{fig:Price_PosRet}.}
		\label{fig:beta1}
\end{figure}

Another disturbing feature of the marginal indifference price is that it is essentially independent of the risk aversion. That it is completely independent of the risk aversion $\alpha$ in the case of exponential utility is obvious from \eqref{eq:PricingPIDEExp}. But also for power utility, it is independent of $\beta$ in the limit $\beta \to \infty$. Using \eqref{eq:betaLimit} and comparing \eqref{eq:PricingPIDEPower} and \eqref{eq:PricingPIDEExp}, one easily sees that the marginal indifference price for power utility converges to the one for exponential utility in the limit $\beta \to \infty$. This property was proven in a general setting in \cite{Nutz10}. In our example, this is shown in Figure~\ref{fig:beta1}: The price changes very little with the risk aversion and approaches the price for exponential utility in the limit $\beta \to \infty$.

\subsection{The implied drift}
The two features just discussed, the drift dependence and the essential risk aversion independence of the marginal indifference price seem to limit the practical applicability of the framework. We also note that the essence of the indifference price is that it takes into account how well the option trade matches to the optimal investment strategy. But typically the investment strategy a bank chooses is not derived from the model that is used to price options. A possible way out is a change of perspective: One takes the actual investment strategy as given and tries to take it into account for the valuation and hedging of options. This is possible straightforwardly, as \eqref{eq:PricingPIDEPower}, \eqref{eq:PricingPIDEExp}, \eqref{eq:OptimalMarginalHedgePower} and \eqref{eq:OptimalMarginalHedgeExp} do not contain the original drift directly, but only via $\tilde \pi^*$ or $\bar \pi^*$. In the case of power utility one would thus set $\tilde \pi^*$ to the fraction of the wealth that is actually invested in the asset and use \eqref{eq:PricingPIDEPower} and \eqref{eq:OptimalMarginalHedgePower}. In the case of exponential utility, one uses the actual amount invested in the asset and the risk aversion $\alpha$ to compute $\bar \pi^*$ via \eqref{eq:barPi}. It is easily seen that this amounts to a change of the drift in the original problem. One may thus speak of an implied drift. Note however, that this implied drift need not be computed for pricing and hedging. It suffices to know the actual investment strategy.

\begin{figure}
	\centering
		\includegraphics[scale = 0.5]{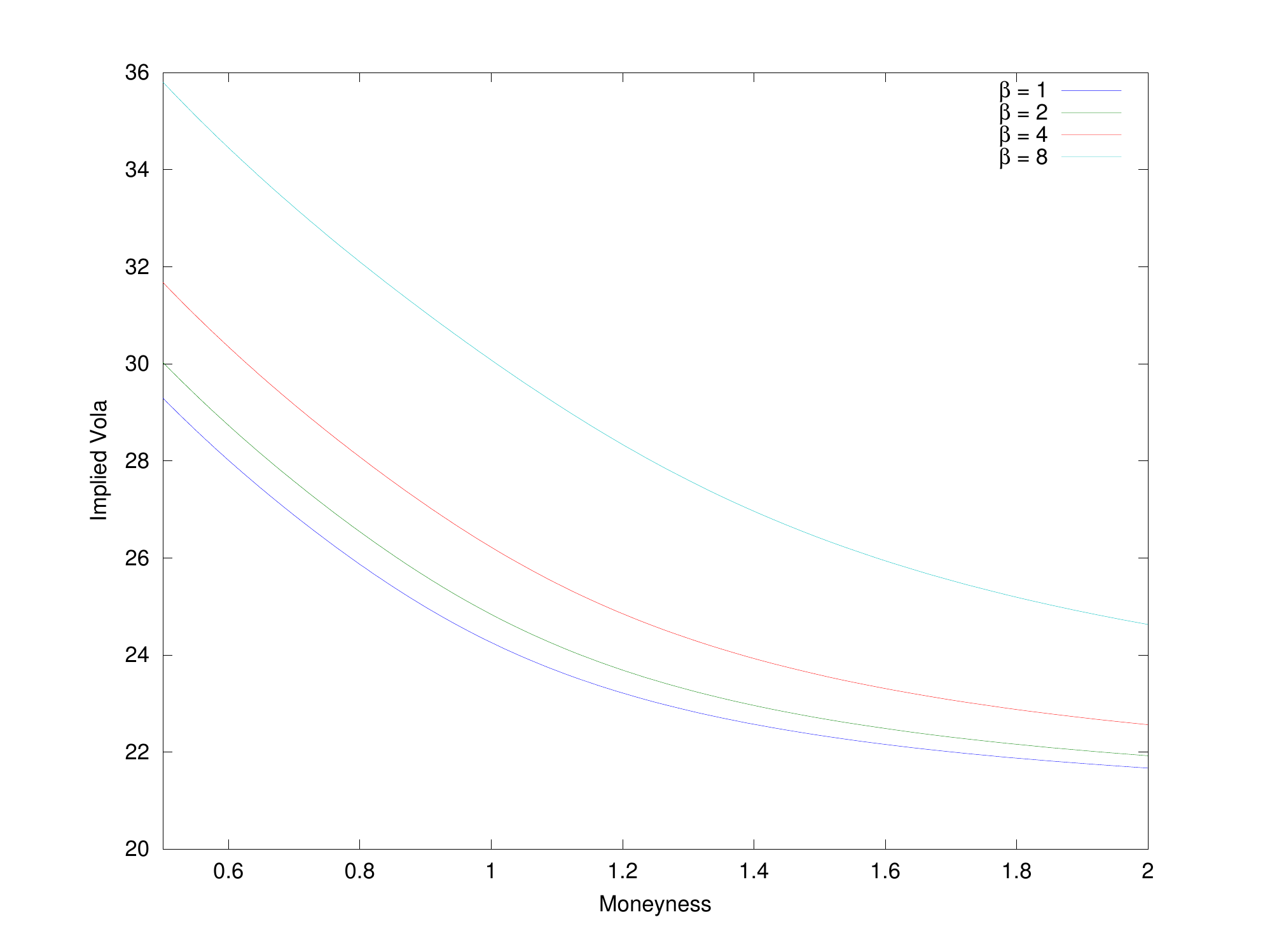}
	\caption{Implied volatilities of the marginal indifference prices obtained for different risk aversions for the parameters $\tilde J= -0.25$, $\lambda = 0.25$, $\sigma = 0.2$, $r=0$, $T = 1$, $\tilde \pi^* = 0.5$. Note that the latter implicitly defines a drift, which is different in the four cases (see text).}
		\label{fig:beta2}
\end{figure}

This change of perspective solves the problems discussed above: One does not need to know the drift, and the price and hedge will in general depend on the risk preference. This is exemplified in Figure~\ref{fig:beta2}. We see that the marginal indifference price increases considerably with the risk aversion. We note however, that also the opposite effect is possible: For a negative actual, i.e., optimal, investment strategy, the marginal indifference price decreases with risk aversion. Again, this is due to the fact that by selling a put the investor can hedge the risk of no jumps happening, to which she is exposed by her investment strategy. Nevertheless, the marginal indifference price is always greater than the Black--Scholes price, in which the jump component is neglected. Finally, we note that for $\pi^* = 0$, one recovers Merton's price. This, however, is not true for the marginal optimal hedge, which coincides with the minimal variance hedge \eqref{eq:mvHedge} for $\alpha = 0$ in that case.

\subsection{The hedge}

\begin{figure}
	\centering
		\includegraphics[scale = 0.5]{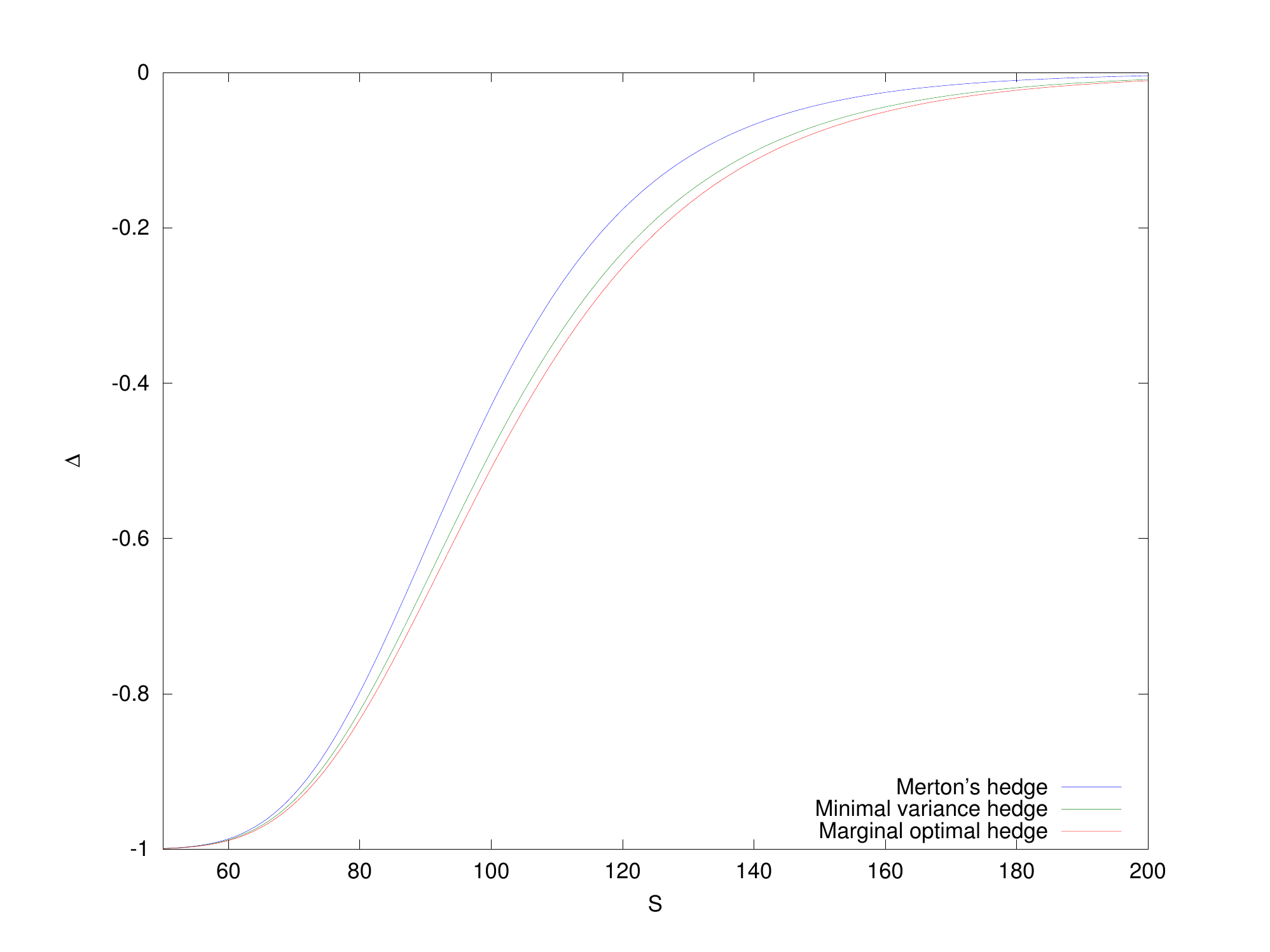}
	\caption{The different hedging strategies (expressed in units of the asset) for a put with strike $K=100$ using the same parameters as in Figure \ref{fig:Price_PosRet}.}
		\label{fig:Delta_PosRet}
\end{figure}

\begin{figure}
	\centering
		\includegraphics[scale = 0.5]{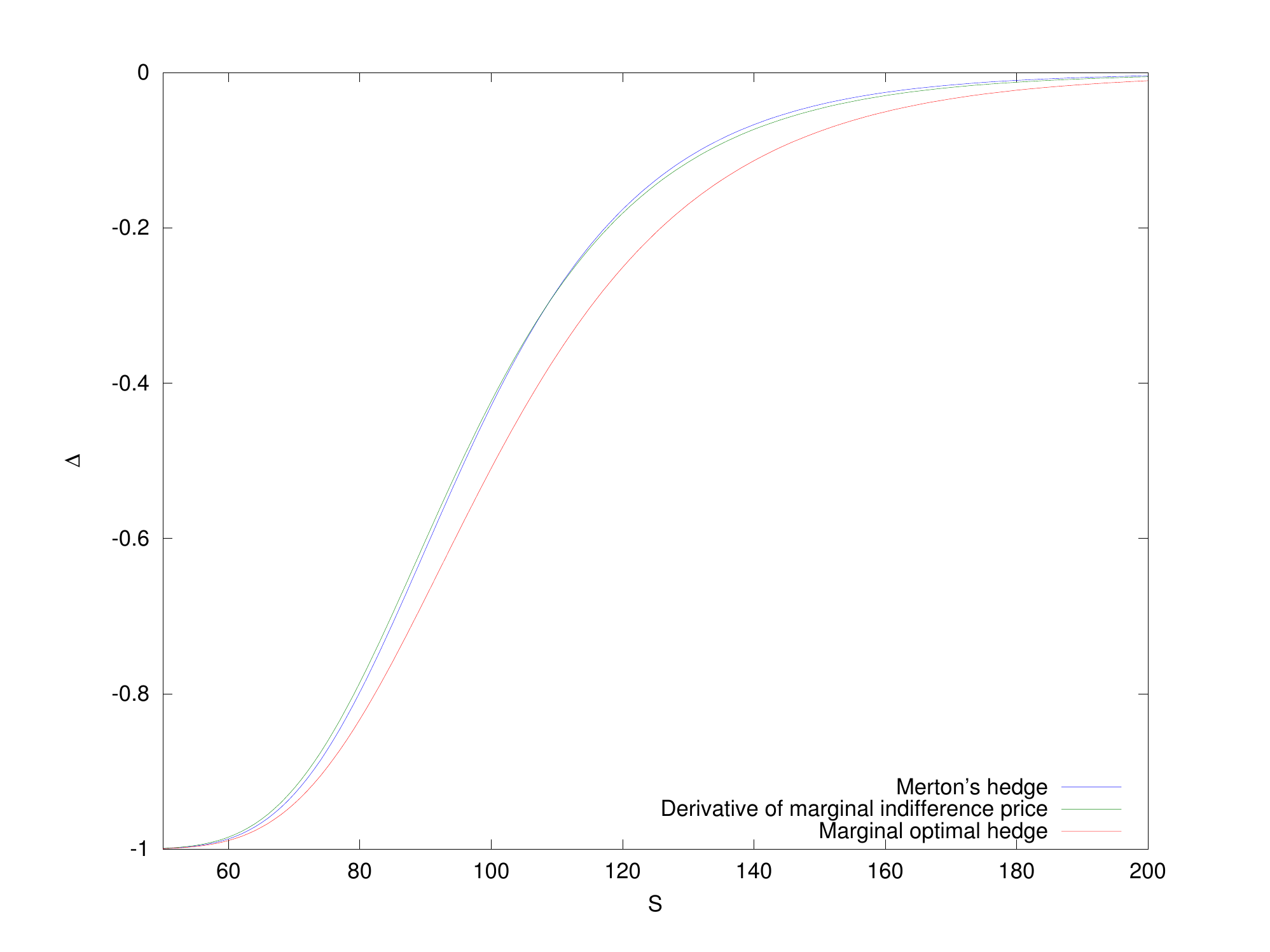}
	\caption{Comparison of Merton's hedge, the marginal optimal hedge and the derivative of the marginal indifference price for the same parameters as in Figure~\ref{fig:Delta_PosRet}.}
		\label{fig:Delta_PosRet_2}
\end{figure}

We now discuss the hedges corresponding to the prices considered before. Figure~\ref{fig:Delta_PosRet} shows the hedges for the prices shown in Figure~\ref{fig:Price_PosRet}. While the minimal variance and the marginal optimal hedge are relatively close together, the deviation from Merton's hedge is noticeable. Heavily out of the money ($S=200$), the relative difference is over 150\%. Note that this strong deviation stems mainly from the new hedging formula \eqref{eq:OptimalMarginalHedgePower} and not so much from using a different price. This can be seen from Figure \ref{fig:Delta_PosRet_2} where, for the same parameters as above, Merton's hedge and the optimal marginal hedge are compared to the derivative \wrt $s$ of the marginal indifference price. This derivative is quite close to Merton's hedge, so for hedging purposes it seems to be more important to use the appropriate hedging formula than to use the correct price.

\begin{figure}
	\centering
		\includegraphics[scale = 0.5]{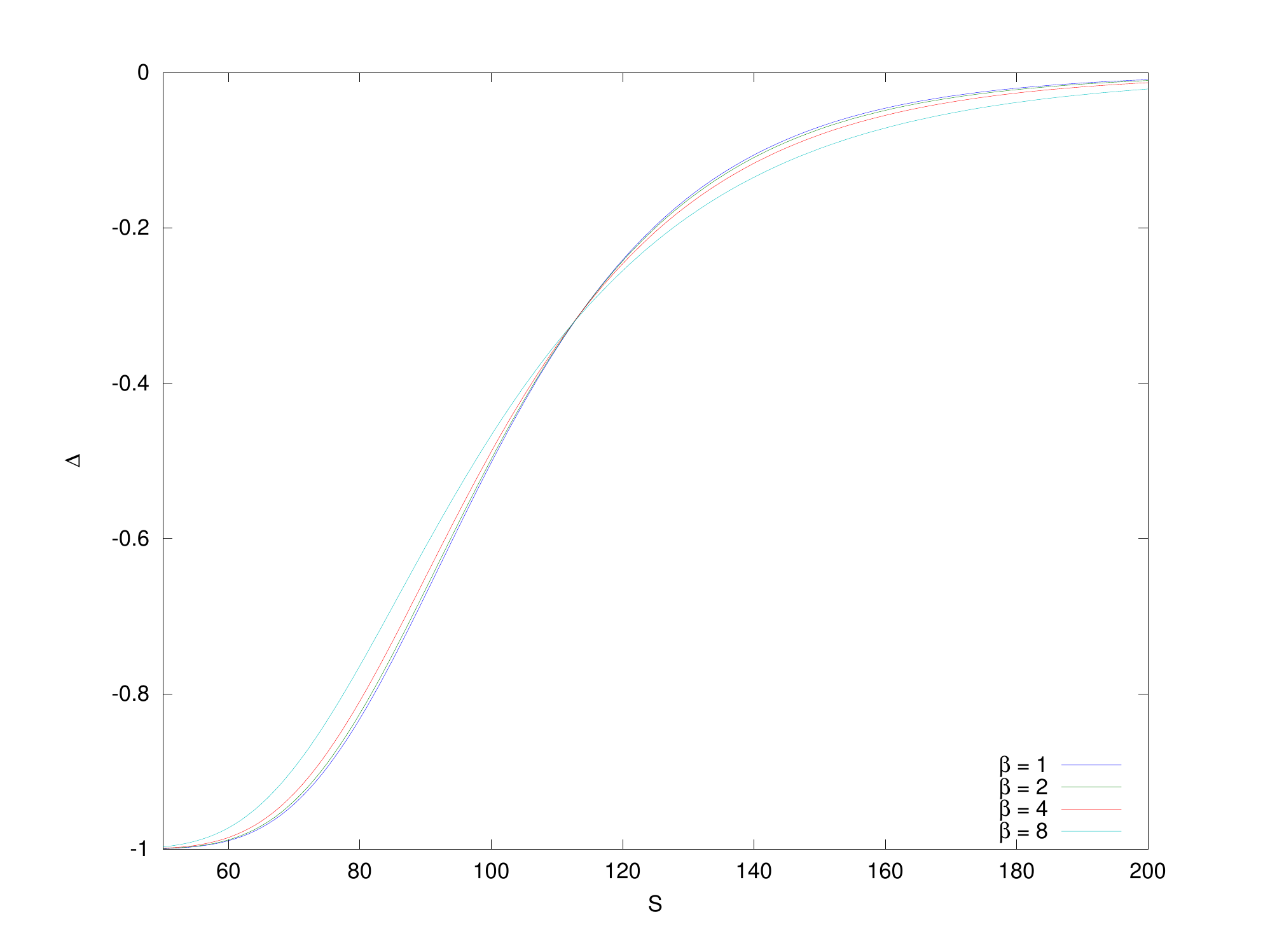}
	\caption{The marginal optimal hedging strategies corresponding to the prices plotted in Figure~\ref{fig:beta2}.}
		\label{fig:beta3}
\end{figure}

Finally, we compare the marginal optimal hedges corresponding to the prices shown in Figure~\ref{fig:beta2}. These are shown in Figure~\ref{fig:beta3}. We see the expected behavior, i.e., for out of the money puts the higher the risk aversion the shorter the investors are in the asset in order to hedge against downward jumps.

\section{Summary \& Outlook}
We discussed marginal utility based pricing and hedging for the case of a jump diffusion process. We pointed out two problems that seem to limit the practical applicability of the framework: The drift dependence and the essential risk aversion independence of the marginal indifference price and the corresponding hedge. We proposed to circumvent these by a change of perspective, by interpreting the actual investment strategy as the optimal one. We also compared the marginal utility based framework conceptually and concretely in a toy model with the minimal variance and Merton's framework.

It would be desirable to apply the framework to more realistic models like a log-normal jump distribution or variance-gamma processes. While this is no problem in principle, we note that by the inclusion of a risk preference, the jump distribution is changed. Thus, computational methods that rely on a particular form of the jump distribution may no longer be applicable.


\

{\bf \noindent Acknowledgements}
\noindent This work is based on a dissertation for the part-time MSc in Mathematical Finance at Oxford University, which was written under the supervision of Jan Obloj. It is a pleasure to thank him for his support and encouragement. I am also grateful to d-fine GmbH, Frankfurt a.~M., Germany, for making my studies in Oxford possible.

\end{document}